# Investigation of Vortex Structures in Gas-Discharge Nonneutral Electron Plasma: III. Pulse Ejection of Electrons at the Formation and Radial Oscillations of Vortex Structure

## N. A. Kervalishvili

Iv. Javakhishvili Tbilisi State University, E. Andronikashvili Institute of Physics,
6 Tamarashvili Str., Tbilisi 0186, Georgia.   <n_kerv@yahoo.com>

**Abstract.** The results of experimental investigations of electron ejection from gas-discharge nonneutral electron plasma at the formation and radial oscillations of vortex structure have been presented. The electrons are injected from the vortex structure and the adjacent region of electron sheath in the form of pulses the duration and periodicity of which are determined by the processes of evolution and dynamics of this structure. The possible mechanisms of pulse ejection of electrons are considered. The influence of electron ejection on other processes in discharge electron sheath is analyzed.

**I. Introduction**

In [1] the experimental methods of investigation developed for studying the local inhomogeneities in gas-discharge nonneutral electron plasma were described. These methods allowed to detect the solitary vortex structures, to investigate their properties, dynamics, and the processes of their formation and evolution [2]. The vortex structures are the inhomogeneities stretched along the magnetic field the density of which is much more than the density of the electron background surrounding them. The vortex structures are formed in discharge electron sheath at the origination of diocotron instability and exist during the time much more than the time of electron-neutral collisions. They play an important role in all processes taking place in the discharge electron sheath. One of such processes is the ejection of electrons from the sheath of gas-discharge nonneutral electron plasma to the end cathodes of discharge device. The average value of the current of these electrons is sufficiently large and reaches 50% of the value of discharge current [3-4].

The electron current on the end cathodes was first observed more than 50 years ago at the investigation of the low pressure discharge in crossed electric and magnetic fields in the geometry of inverted magnetron [5]. Further, it was shown that this current is ejected in the form of periodically following pulses with the frequency proportional to the pressure of neutral gas. [6, 7]. However, before detection of vortex structures the mechanism of formation of such electrons remained unknown. At the investigation of vortex structures in gas-discharge nonneutral electron plasma it was observed that the formation, the radial oscillations and the approach of vortex structures are accompanied by pulse ejection of electrons from electron plasma to the end cathodes [8-11]. The analysis of the experimental results allowed to investigate this process in detail and to explain the mechanism of electron ejection.

The processes of formation and dynamics of vortex structures are described in detail in [2]. In the present work, the ejection of electrons from vortex structures and adjacent regions of electron sheath accompanying these processes is studied. In paragraph II the process of pulse electron ejection at the formation of vortex structure is investigated. In paragraph III the process of pulse ejection of electrons at radial oscillations of vortex structure is studied. In paragraph IV the mechanisms of pulse electron ejection are considered. In paragraph V the influence of electron ejection on other processes in gas-discharge nonneutral electron plasma is discussed.

**II. Ejection of electrons at formation of vortex structure**

Fig 1 presents the oscillograms showing the ejection of electrons at formation of quasi-stable vortex structure in the geometry of inverted magnetron. Upper oscillograms are the oscillations of electric field on the anode wall probe, and the lower – the current of electrons on the end cathodes.

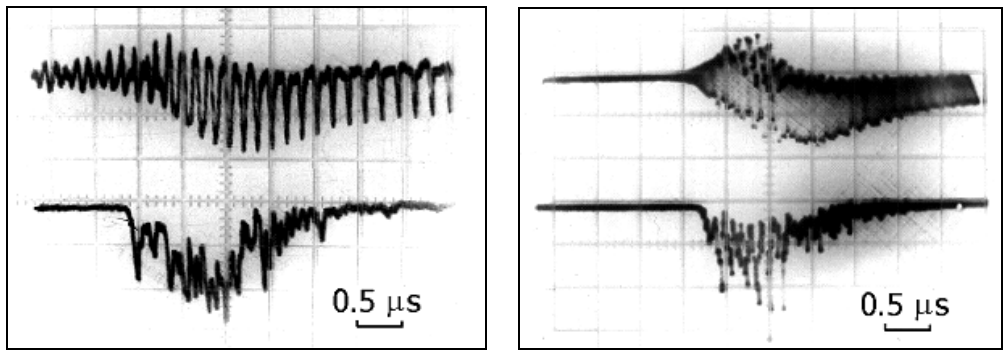

Fig.1 Ejection of electrons at formation of vortex structure
Left: $r_a = 2.0 cm$; $r_c = 3.2 cm$; $L = 7 cm$; $B = 1.5 kG$; $V = 1.0 kV$; $p = 1 \times 10^{-5} Torr$
Right: $r_a = 1.0 cm$; $r_c = 3.2 cm$; $L = 7 cm$; $B = 1.5 kG$; $V = 1.0 kV$; $p = 1 \times 10^{-5} Torr$

Let us consider the process of electron ejection in detail. The pulse of electron ejection at formation of quasi-stable vortex structure lasts considerably longer than the period of rotation of vortex structure about the axis of discharge device. It starts at the moment of formation of a hole in electron sheath and lasts until the quasi-stable vortex structure is formed completely. The pulse has a complex shape reflecting the regular and irregular processes taking place during the formation of vortex structure. The most typical among these processes is described in detail in [2,12]. This is the fragmentation of sheath and the merging of separate fragments. Each of these processes is accompanied by additional narrow pulses of electron ejection which are imposed on the main pulse connected with the bunching of electrons in a compact dense structure. The examples of such narrow pulses are given on the oscillograms in figs. 2 and 3. Here, as in Fig. 1, the upper oscillogram is the oscillations of electric field on the anode wall probe, and the lower – the current of electrons on the end cathodes.

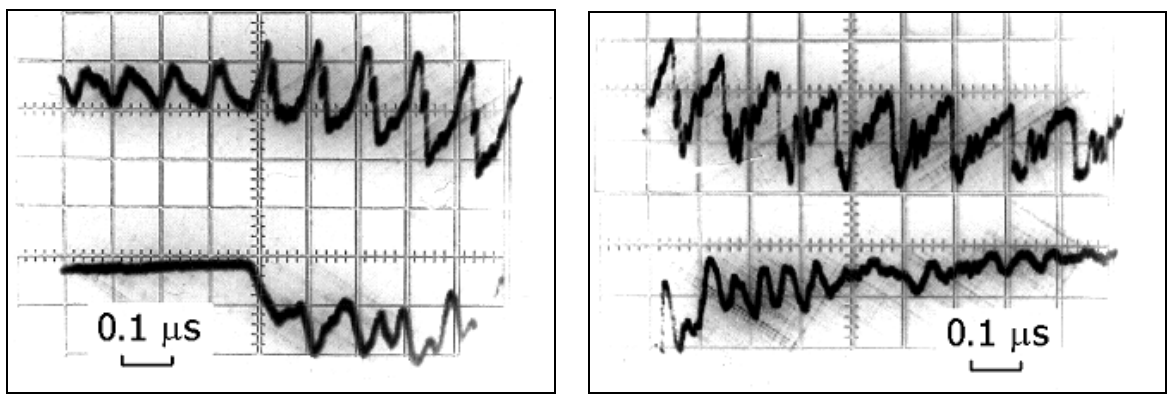

Fig.2 Fragments of electron ejection at formation of vortex structure
$r_a = 2.0 cm$; $r_c = 3.2 cm$; $L = 7 cm$; $B = 1.5 kG$; $V = 1.0 kV$; $p = 2 \times 10^{-5} Torr$

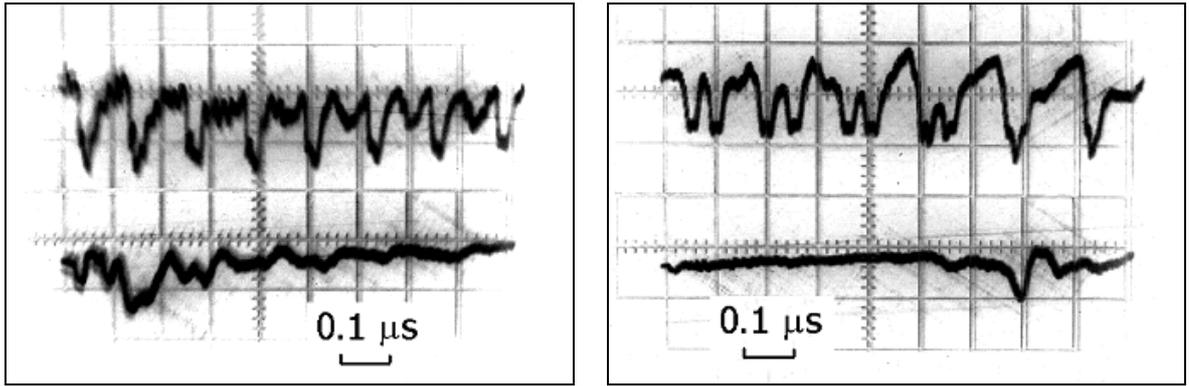

Рис.3 Fragments of electron ejection at formation of vortex structure
$r_a = 2.0 cm$; $r_c = 3.2 cm$; $L = 7 cm$; $B = 1.5 kG$; $V = 1.0 kV$; $p = 1 \times 10^{-5} Torr$

The left oscillograms in Fig. 2 show that the pulse of electron ejection at formation of quasi-stable vortex structure starts from the moment of formation of a hole in discharge electron sheath. (The appearance of a hole is indicated by the increase and the narrowing of the positive half-period of oscillations of electric field on the anode wall probe [2]). From the moment of appearance of a hole there starts the process of bunching the electrons in a clump at the point of electron sheath diametrically opposite to the hole. Simultaneously, the fragmentation of the "tail" of the formed vortex structure takes place (upper oscillogram in Fig. 2 on the right). This process is accompanied by narrow pulses of electron ejection. From the fragments the second vortex structure is formed (upper oscillogram in Fig. 3 on the left). Further, the approach and then the merging of both vortex structures take place (upper oscillogram in Fig.3 on the right). The formation and the merging of vortex structures are accompanied by narrow pulses of electron ejection. At this point the process of formation of quasi-stable vortex structure is completed and, correspondingly, the electron ejection accompanying this process is finished (Fig. 1).

The oscillograms presented in Figs. 2 and 3 show the typical processes taking place at formation of quasi-stable vortex structure. However, at each following formation of vortex structure, the duration and the intensity of these separate processes, the number of fragments, etc. can differ strongly indicating their irregularity. At the same time, the resulted averaged pattern of periodical formation of quasi-stable vortex structure seems quite regular, as it is seen in Fig. 4. Here, as in Figs. 1-3, the upper oscillogram is the oscillations of electric field on the anode wall probe, and the lower – the current of electrons on the end cathodes.

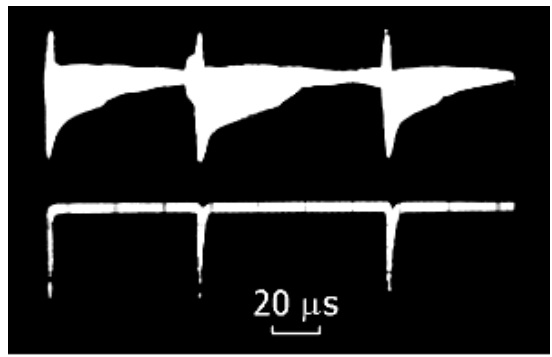

Fig. 4 Periodical formation of quasi-stable vortex structure
$r_a = 2.0 cm$; $r_c = 3.2 cm$; $L = 7 cm$; $B = 1.5 kG$; $V = 1.5 kV$; $p = 1 \times 10^{-5} Torr$

## III. Ejection of electrons at radial oscillations of vortex structure

Other source of pulse ejection of electrons is the radial oscillations of vortex structure taking place during the orbital instability observed first in [8] and described in detail in [2]. Fig. 5 shows the oscillograms of oscillations of electric fields on wall probes of anode (upper) and cathode (lower) in the case of periodical appearance of orbital instability (left) and directly during the orbital instability (right) in magnetron geometry of discharge device.

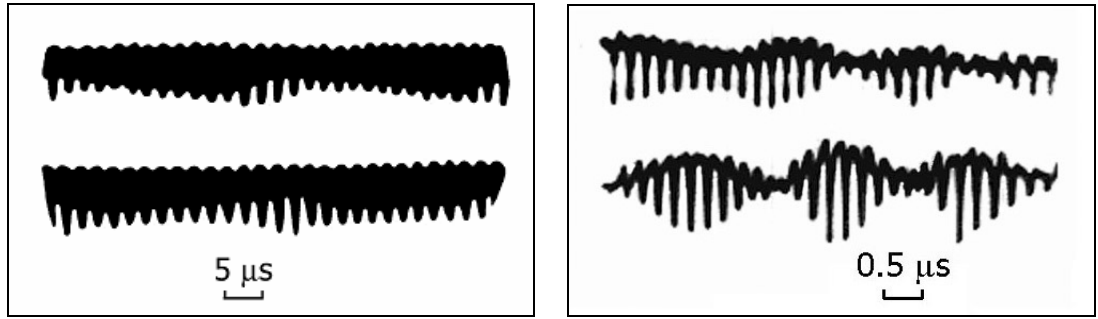

Fig. 5 Orbital instability in magnetron
$r_a = 3.2 cm$; $r_c = 1.0 cm$; $L = 7 cm$; $B = 1.2 kG$; $V = 1.5 kV$; $p = 5 \times 10^{-6} Torr$

In Figs. 6 and 7 the ejection of electrons is shown during the orbital instability. Lower oscillograms show the oscillations of electric field on the cathode wall probes, and the upper oscillograms - the current of electrons on the end cathodes.

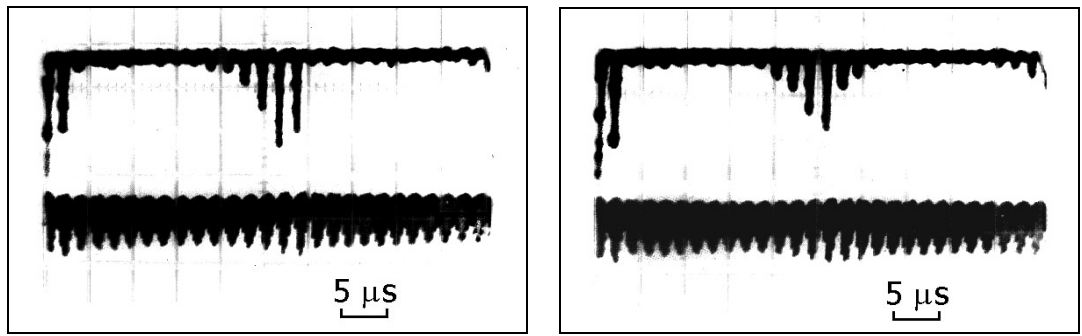

Fig. 6. Ejection of electrons during the orbital instability in magnetron
$r_a = 3.2 cm$; $r_c = 1.0 cm$; $L = 7 cm$; $B = 1.2 kG$; $V = 1.5 kV$; $p = 5 \times 10^{-6} Torr$

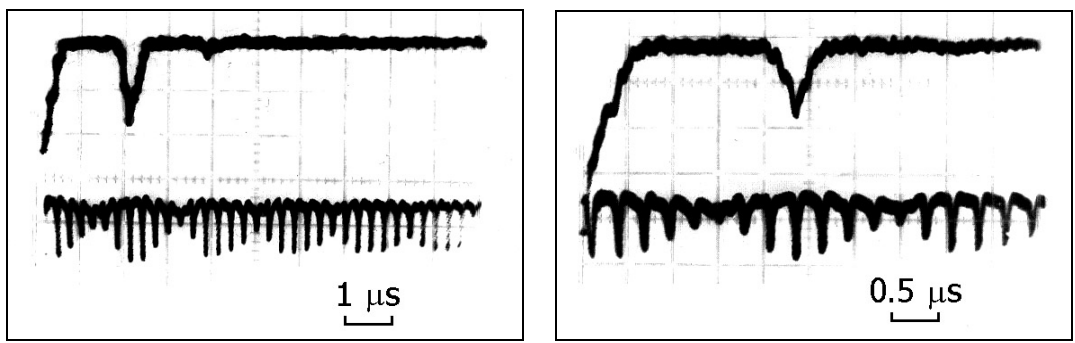

Fig. 7. Ejection of electrons during the orbital instability in magnetron
$r_a = 3.2 cm$; $r_c = 1.0 cm$; $L = 7 cm$; $B = 1.2 kG$; $V = 1.5 kV$; $p = 5 \times 10^{-6} Torr$

As it follows from Figs. 5-7, the drift trajectory of vortex structure is a flat spiral and the ejection of electrons takes place in the process of gradual approach of vortex structure to the cylindrical cathode. Therefore, each pulse of electron ejection lasts during several rotations of vortex structure about the axis of discharge device (Fig 7). In the process of approaching the vortex structure to the anode, its motion takes place as well in a spiral, however, the ejection of electrons is absent. At appearing the orbital instability the amplitude of each following pulse of electron current increases (Fig. 6). This is correlated with more and more approach of vortex structure to the cathode at its each following radial oscillation. This is evidenced by the increase of each following amplitude of oscillations of electric field on the cathode wall probe and, correspondingly, by the decrease of the amplitude of oscillations of electric field on the anode wall probe. Then, the amplitude of pulses of electron current and of radial oscillations of vortex structure decreases. The amplitude of pulses of electron current decreases more rapidly than the amplitude of radial oscillations of vortex structure.

In Penning cell the orbital instability and the accompanying pulses of electron current in the most cases behave in the same way as in magnetron geometry. Fig. 8 presents the oscillograms of oscillations of electric field on the anode wall probe (upper) and of electron current on the end cathodes (lower) in Penning cell. As it is seen from the figure, the amplitudes of pulses of electron current and of radial oscillations of vortex structure decrease gradually

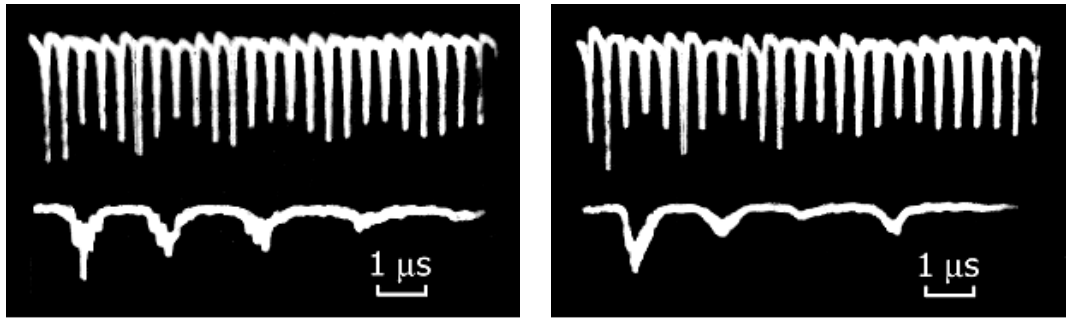

Fig. 8 Ejection of electrons during the orbital instability in Penning cell
$r_a = 3.2 cm$; $L = 7 cm$; $B = 2.0 kG$; $V = 2.0 kV$; $p = 5 \times 10^{-6} Torr$

However, in some cases, the amplitude of oscillations of electric field on the anode wall probe decreases very strongly and it is accompanied by the intense ejection of electrons. This indicates the fact that sometimes the vortex structure approaches the axis of Penning cell quite close causing great losses of electrons of vortex structure. Fig.9 presents the oscillograms showing such behavior of vortex structures. Here, the upper oscillogram is the oscillations of electric field on the anode wall probe, and the lower – electron current on the end cathodes.

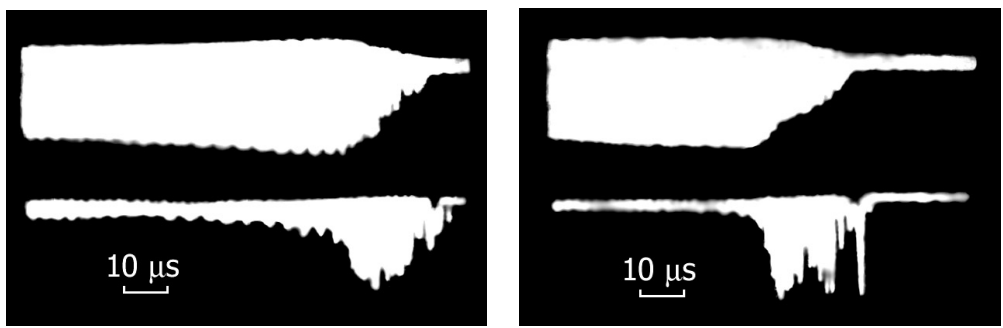

Fig. 9 Approach of vortex structure to the axis of Penning cell
$r_a = 3.2 cm$; $L = 7 cm$; $B = 2.0 kG$; $V = 2.0 kV$; $p = 5 \times 10^{-6} Torr$

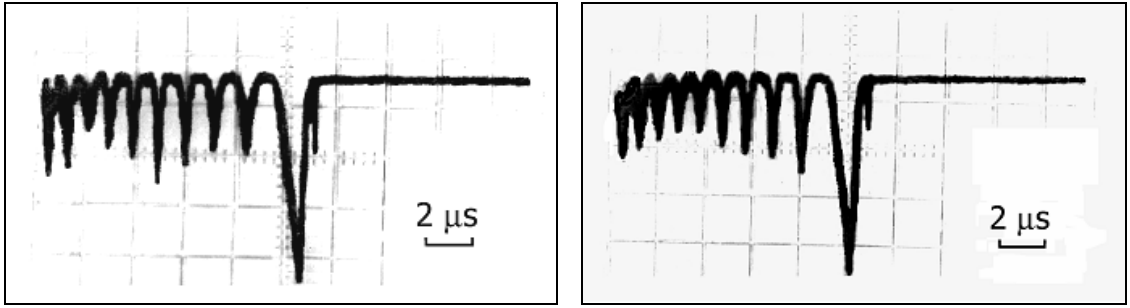

Fig.10 Ejection of electrons at the approach of vortex structure to the axis of Penning cell
$r_a = 3.2 cm$; $L = 7 cm$; $B = 1.5 kG$; $V = 2.0 kV$; $p = 5 \times 10^{-6} Torr$

Fig. 10 shows the pulses of electron ejection in more large scale. The last pulse is the greatest. Just it is related to the approach of vortex structure to the axis of Penning cell. To the right of this pulse a small narrow pulse is located. The origin of this pulse is explained by Fig. 11.

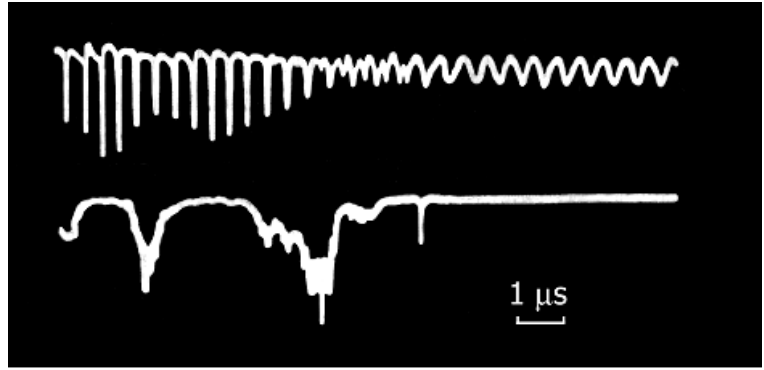

Fig. 11 Short-lived second vortex structure in Penning cell
$r_a = 3.2 cm$; $L = 7 cm$; $B = 2.0 kG$; $V = 2.0 kV$; $p = 5 \times 10^{-6} Torr$

The upper oscillogram in this figure is the oscillations of electric field on the anode wall probe, and the lower – of electron current on the end cathodes. On the oscillograms it is seen that the approach of vortex structure to the axis of Penning cell and a great loss of electrons of vortex structure are accompanied by the origination of the second vortex structure and then by its merging with the initial vortex structure. The both processes (origination and merging) lead to the appearance of narrow pulses of electron ejection in the similar way as at the formation of quasi-stable vortex structure in inverted magnetron (Fig. 3).

**IV. Mechanism of pulse ejection of electrons**

As it follows from the experimental results, the ejection of electrons from the gas-discharge nonneutral electron plasma is always connected with the vortex structures [8-11]. Above, we described in detail the pulse ejection of electrons at the formation of vortex structure and at its radial oscillations, more exactly, at its radial displacement to the cylindrical cathode. (Ejection of electrons taking place at the interaction of vortex structures [8], as well as in the case of one stable vortex structure will be considered later on in detail). It should be noted that the ejection of electrons always takes place from the region of electron sheath where the vortex structure is located. If the duration of ejection is comparable or more than the time of turn of vortex structure about the axis of discharge device, the region of ejection moves together with the vortex structure about the axis of discharge device. This is seen well in Fig. 12 showing the oscillograms of oscillations of electric field on the anode wall probe (upper) and of electron ejection current through the narrow radial slit on the end cathode (lower).

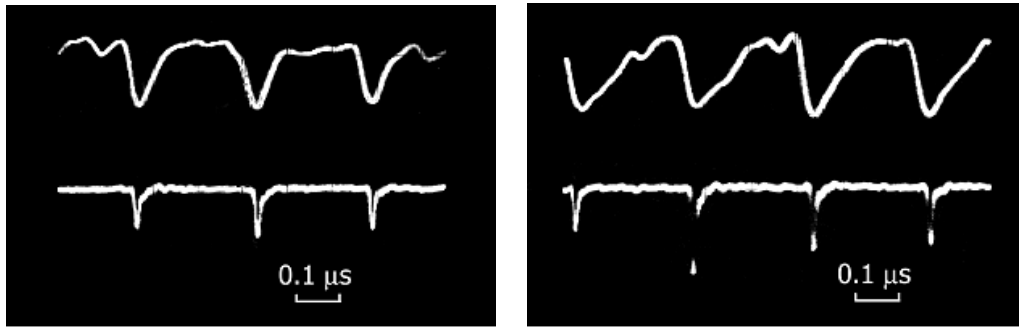

Fig. 12 Electron current through the narrow radial slit on the end cathode
$r_a = 3.2 cm$; $L = 7 cm$; $B = 1.9 kG$; $V = 1.0 kV$; $p = 1 \times 10^{-5} Torr$

The slit is located on the same azimuth as the anode and cathode wall probes. As it is seen from the figure, the region of electron ejection coincides with the place of existence of vortex structure and rotates together with it about the axis of discharge device. From this it follows that the ejection of electrons takes place from the vortex structure and from the adjacent region of electron sheath. As the density of electrons of vortex structure is much more than the density of electron sheath, and the width of slit is much less than the transverse dimensions of vortex structure (narrow slit), the width of ejection pulse through the slit can be identified with the azimuth dimension of vortex structure.

Now let us consider the possible mechanism of electron ejection. For nonneutral electron plasma the value of potential barrier along the magnetic field depends on the distance from the anode surface and on the average density of electrons in this region of sheath. From this it can be supposed that the ejection of electrons from the vortex structure takes place at the decrease of potential barrier along the magnetic field or at the expense of local increase of electron density at the formation of vortex structure, or at the displacement of the vortex structure itself to the region with lower potential barrier during the orbital instability. However, the fact of the decrease of potential barrier itself is not enough for ejection of electrons. In the discharge the electrons acquire the transverse velocity at the expense of electric field, and longitudinal velocity appears as a result of electron-neutral collisions. The nearer is to the anode and more is the density of electron sheath the more is the value of electric field and more is the average transverse velocity of electrons. Correspondingly, the more is the longitudinal velocity acquired as a result of electron-neutron collisions. At the same time, the longitudinal velocity acquired by electrons does not depend anymore on radial displacement of electrons, it does not depend either on the change of potential at the point of existence of electrons. From this it follows that the electrons that acquired the longitudinal velocity at the expense of collisions with neutral atoms at high potential barrier and, correspondingly, at high average energy of electrons, leave the electron sheath at lower potential barrier if the longitudinal energy acquired by them earlier appears to be enough to overcome this decreased potential barrier. It should be noted that the potential barrier is not the same for all parts of vortex structure. For the part of vortex structure that is located closer to the cylindrical cathode, the potential barrier is less. As the vortex structure rotates about its own axis and the angular velocity of this rotation is much more than the angular velocity of the rotation of vortex structure about the axis of discharge device all parts of vortex structure appear sequentially in the region of potential barrier being more "favorable" for ejection of electrons. Thus, the rotation of vortex structure will assist the ejection of electrons.

The assumption that the electrons acquire the necessary longitudinal velocity at the expense of electron-neutral collisions is justified by the fact that the time between periodical ejections of electrons connected with the formation of vortex structure in inverted magnetron or with the orbital instability in magnetron and Penning cell is of the order of time tens of electron-neutral collisions [2]. During the orbital instability the time between the following pulses of electron ejection, on the contrary, is much less than the time of electron-neutral collisions. However, at each following radial displacement the vortex structure moves more and more away from the anode surface appearing in the region of more and more low potential barrier and, therefore, the amplitude of electron pulses continues to increase. As soon as the amplitude of radial oscillations of vortex structure stops to increase, the amplitude of electron pulses begins to decrease as the number of electrons having sufficient longitudinal velocity decreases rapidly.

**V. Discussion and conclusion**

The experiments described above were carried out in three geometries of discharge device – magnetron, inverted magnetron and Penning cell – and in all cases the solitary vortex structures were responsible for the ejection of electrons from electron sheath to the end cathodes. Hence, the vortex structures, which practically always present in gas-discharge nonneutral electron plasma, create a new mechanism of electron losses. This mechanism is connected, on the first hand, with electron-neutral collisions and, on the other hand, with collisionless processes – formation, rotation, interaction and dynamics of vortex structures. As soon as the "longitudinal energy" of the electron appears above the potential barrier, the electron leaves the sheath or the vortex structure practically instantly. Thanks to such "rapidity" this mechanism begins to play the defining role in the process of limiting the density of electron sheath. This means that the density of electron sheath will be determined not by the balance between the ionization and the mobility of electrons across the magnetic field, as it was assumed earlier, but by the "critical" electron density, at which there appears the diocotron instability and the solitary vortex structures are formed. In [13, 14] the model of electron sheath was considered, in which the equilibrium density of electrons is determined by "critical" electron density, at which there appears the diocotron instability. The model describes well the current characteristics of gas-discharge nonneutral electron plasma both in magnetron geometry and in the geometry of inverted magnetron, including the case when the magnetic field is not parallel to the anode surface [14]. Thus, the agreement of the model with the experimental results confirms the significant role of electron ejection in formation and evolution of discharge electron sheath.

The considered mechanisms of pulse ejection of electrons describe well the qualitative pattern of the process of ejection from vortex structure and adjacent regions of electron sheath. However, for quantitative description of electron ejection the detailed calculations are necessary. In [15] the numerical simulation of low-pressure discharge in magnetron geometry of discharge device was made at about the same geometrical and electrical parameters of the discharge, as in the experiments described above. At simulation the real physical processes in the discharge were taken into account, in the first place, the ionization and the possible losses of electrons on end cathodes, if their longitudinal velocity allowed to overcome the potential barrier. Though the obtained results differ slightly from the experimentally observed pattern, they demonstrate quite clearly the periodic transformation of electron sheath into self-organizing vortex structures and the pulse ejection of electrons on end cathodes at their approach, merging and the approach to the cylindrical cathode.

Let us note one more physical process in discharge electron sheath connected with the vortex structures. During the orbital instability the vortex structure makes the periodical radial displacements from the anode to the cylindrical cathode and back. Approaching the cathode, the vortex structure loses a part of the electrons acquired by it being close to the anode. Thus, the vortex structure transfers the electrons to the part of electron sheath, where the potential barrier is less and, hence, the probability of going the electrons away along the magnetic field is higher. Consequently, the vortex structure makes something like the convective transport of electrons across the magnetic field from the region of sheath located close to the anode to the region of sheath located close to the cathode, i.e., to the reverse direction - opposite the electric field forces.

**References**

[1] N.A. Kervalishvili, Investigation of vortex structures in gas-discharge nonneutral electron plasma: I. Experimental technique, arXiv:1502.02516 [physics.plasm-ph] (2015).

[2] N.A. Kervalishvili, Investigation of vortex structures in gas-discharge nonneutral electron plasma: II. Vortex formation, evolution and dynamics, arXiv:1502.07945 [physics.plasm-ph] (2015).

[3] N.A. Kervalishvili, Effect of anode orientation on the characteristics of a low-pressure discharge in a transverse magnetic field, Sov. Phys. Tech. Phys. **13**, 476 (1968) [Zh. Tekh. Fiz. **38**, 637 (1968)].

[4] N.A. Kervalishvili and V.P. Kortkhonjia, Low-pressure discharge in a transverse magnetic field, Sov. Phys. Tech. Phys. **18**, 1203 (1974) [Zh. Tekh. Fiz. **43**, 1905 (1973)].

[5] J.P. Hobson and P.A. Redhead, Operation of an inverted-magnetron gauge in the pressure range $10^{-3}$ to $10^{-12}$ mm Hg, Can. J. Phys. **36**, 271 (1958).

[6] N.A. Kervalishvili, Instabilitis of a low-pressure discharge in a transverse magnetic field, Sov. Phys. Tech. Phys. **13**, 580 (1968) [Zh. Tekh. Fiz. **38**, 770 (1968)].

[7] E.M. Barkhudarov, N.A. Kervalishvili and V.P. Kortkhonjia, Anode sheath instability and high-energy electrons in a low-pressure discharge in a transverse magnetic field, Sov. Phys. Tech. Phys. **17** 1526 (1973) [Zh. Tekh. Fiz. **42** 1904 (1972)].

[8] N.A. Kervalishvili, Rotational instability of a nonneutral plasma in crossed fields $E \perp H$ and generation of electrons of anomalously high energy, Sov. J. Plasma Phys. **15**, 98 (1989) [Fizika Plazmy **15**, 174 (1989)].

[9] N.A. Kervalishvili, Rotating regular structures in a nonneutral plasma in crossed electric and magnetic fields, Sov. J. Plasma Phys. **15**, 211 (1989) [Fizika Plazmy **15**, 359 (1989)].

[10] N.A. Kervalishvili, Evolution of nonlinear structures in nonneutral plasma in crossed fields $E \perp H$, Sov. J. Plasma Phys. **15**, 436 (1989) [Fizika Plazmy **15**, 753 (1989)].

[11] N.A. Kervalishvili, Electron vortices in a nonneutral plasma in crossed $E \perp H$ fields, Phys. Lett. A **157**, 391 (1991)

[12] N.A. Kervalishvili, Formation and dynamics of vortex structures in pure and gas-discharge nonneutral collisionless electron plasmas, arXiv:1308.5415 [physics.plasm-ph] (2013).

[13] G.N. Kervalishvili, J.I. Javakhishvili, and N.A. Kervalishvili, Diocotron instability in an annular sheath of gas-discharge nonneutral electron plasma, Phys. Lett. A **296**, 289 (2002).

[14] N.A. Kervalishvili and G.N. Kervalishvili, The model of gas-discharge nonneutral electron plasma, arXiv:1311.4397 [physics.plasm-ph] (2013); J. Georgian Geophysical Society **12B**, 105 (2008).

[15] J.P. Boeuf, Rotating structures in low temperature magnetized plasmas – insight from particle simulations, Front. Phys. **2**, 00074 (2014).